\documentclass[aps,prbtwocolumn,superscriptaddress,citeautoscript,amssymb,reprint,showpacs,floatfix,longbibliography]{revtex4-2}
\usepackage[utf8]{inputenc}
\usepackage[T1]{fontenc}
\usepackage{float}
\usepackage{amsmath}
\usepackage[normalem]{ulem}
\usepackage{mathptmx}
\usepackage{mathrsfs}
\usepackage[english]{babel}
\usepackage{graphicx}
\usepackage[usenames,dvipsnames]{color}
\usepackage{natbib}
\usepackage{makecell}
\usepackage[thinspace,thinqspace]{SIunits}
\usepackage{nicefrac}
\newcommand{\CRA}{CeRh$_2$As$_2$}

\newcommand{\To}{$T_{\textrm{0}}$}
\newcommand{\Ho}{$H_{\textrm{0}}$}
\newcommand{\Hcr}{$H_{\textrm{cr}}$}

\begin{document}
\author{K.~Semeniuk}
\thanks{correspondence should be addressed to konstantin.semeniuk@cpfs.mpg.de and elena.hassinger@kit.edu}
\affiliation{Max Planck Institute for Chemical Physics of Solids, 01187 Dresden, Germany}

\author{B.~Schmidt}
\affiliation{Max Planck Institute for Chemical Physics of Solids, 01187 Dresden, Germany}

\author{C.~Marcenat}
\affiliation{Universit\'e Grenoble Alpes, CEA, Grenoble-INP, IRIG, Pheliqs, 38000 Grenoble, France}

\author{M.~Pfeiffer}
\affiliation{Institute for Solid State and Materials Physics, Dresden University of Technology, 01062 Dresden, Germany}

\author{A.~Demuer}
\affiliation{Laboratoire National des Champs Magnétiques Intenses (LNCMI-EMFL), CNRS, Universit\'e Grenoble Alpes, 38042 Grenoble, France}

\author{L.~Behera}
\affiliation{Institute for Solid State and Materials Physics, Dresden University of Technology, 01062 Dresden, Germany}

\author{T.~Klein}
\affiliation{Universit\'e Grenoble Alpes, CNRS, Institut N\'eel, 38000 Grenoble, France}

\author{S.~Khim}
\affiliation{Max Planck Institute for Chemical Physics of Solids, 01187 Dresden, Germany}

\author{E.~Hassinger}
\thanks{correspondence should be addressed to konstantin.semeniuk@cpfs.mpg.de and elena.hassinger@kit.edu}
\affiliation{Institute for Quantum Materials and Technologies, Karlsruhe Institute of Technology, Kaiserstraße 12, 76131 Karlsruhe, Germany}
\affiliation{Max Planck Institute for Chemical Physics of Solids, 01187 Dresden, Germany}

\title{Basal-plane anisotropy of field-induced multipolar order in tetragonal \CRA}

\date{\today}

\begin{abstract}
Unconventional superconductivity in Ce-based Kondo-lattice materials emerges almost exclusively in the vicinity of weak dipolar magnetic orders, while higher multipolar orders are only known to occur in a few Pr-based unconventional superconductors and possibly URu$_2$Si$_2$. The multiphase superconductor \CRA\ appears to be a notable exception from this trend. Showing clear signatures of magnetism, this tetragonal system is suspected to host a concomitant quadrupolar order, which could be causing the strong enhancement of the ordering temperature when a magnetic field is applied perpendicular to the fourfold ($c$) axis of the lattice. In this work, we show that the field-temperature phase diagram of \CRA\ has a remarkable basal-plane anisotropy. This finding supports the scenario of coupled magnetic and multipolar ordering, which may have implications for the pairing mechanism of the superconductivity, and guides the development of the next iteration of theoretical models.
\end{abstract}
\maketitle

\textit{Introduction.}---The Kondo-lattice system \CRA\ is a heavy-fermion superconductor that exhibits a very rare magnetic-field-induced transition between two distinct superconducting (SC) phases~\cite{khim2021}. These phases were proposed to be states of the even and odd parity of the SC order parameter~\cite{landaeta2022,yoshida2012,yoshida2014,fischer2023}, earning the material a significant attention.

\begin{figure}[b]
     \includegraphics[width=\columnwidth]{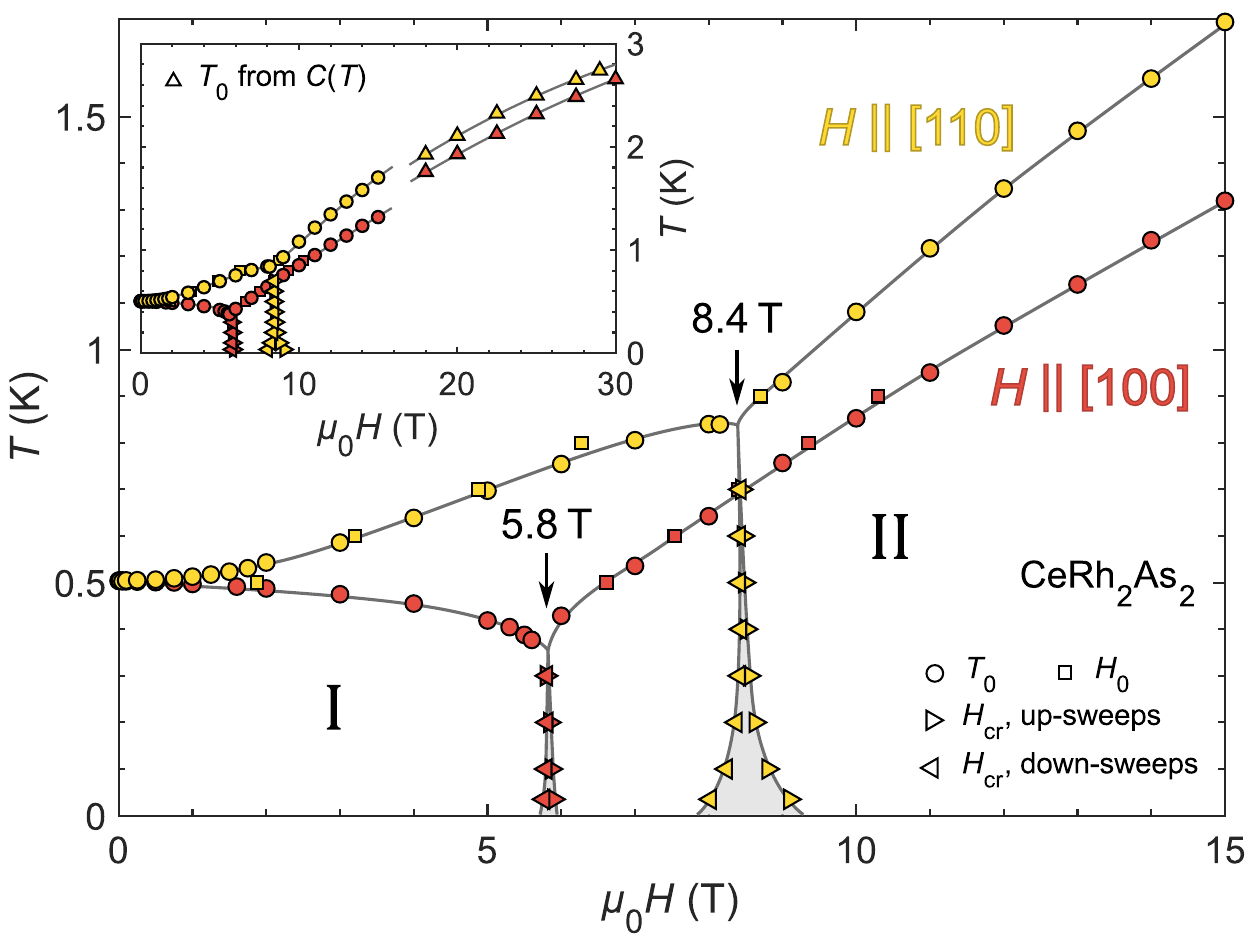}
     \caption{Phase diagrams of \CRA\ as functions of temperature ($T$) and magnetic field ($\mu_{\textrm{0}}H$, where $\mu_{\textrm{0}}$ is the vacuum permeability), applied along the [100] and [110] crystallographic directions. Phases I and II are indicated. The critical temperature (\To) and fields (\Ho, \Hcr) were extracted from measurements of $T$ and $H$ dependence of electrical resistivity, respectively. Solid grey lines are guides for the eye. The hysteretic regions are shaded in light-grey. The superconducting phase is omitted. Inset: an extension of the phase diagrams to higher fields according to heat capacity [$C(T)$] data.}
     \label{fig:PhaseDiagrams}
\end{figure}

However, the multiphase superconductivity is not the only intriguing phenomenon of \CRA. In accordance with the established paradigm of unconventional superconductivity~\cite{mathur1998}, the compound is located close to a pressure-induced quantum critical point of another ordered state, known as phase~I~\cite{hafner2022,semeniuk2023,pfeiffer2024}, which is quite unusual in and of itself. At ambient pressure, phase~I sets in at the critical temperature \To\ (see Fig.~\ref{fig:PhaseDiagrams}), which is highly sensitive to static crystalline defects and reaches 0.54\,K in the cleanest samples obtained so far~\cite{khanenko2025}. The \To\ phase transition has a definite magnetic character, confirmed by muon spin resonance and nuclear magnetic/quadrupole resonance studies, although the ordering vector and moment orientations currently remain unknown~\cite{khim2025,kitagawa2022,kibune2022,ogata2023}. Measurements of inelastic neutron scattering showed a very weak intensity at the (1/2,1/2,0) wave vector~\cite{chen2024}, but the same study could not detect any long-range magnetic order, placing an upper limit of the ordered moment at 0.2 of Bohr magneton.

\begin{figure}[t]
    \includegraphics[width=\columnwidth]{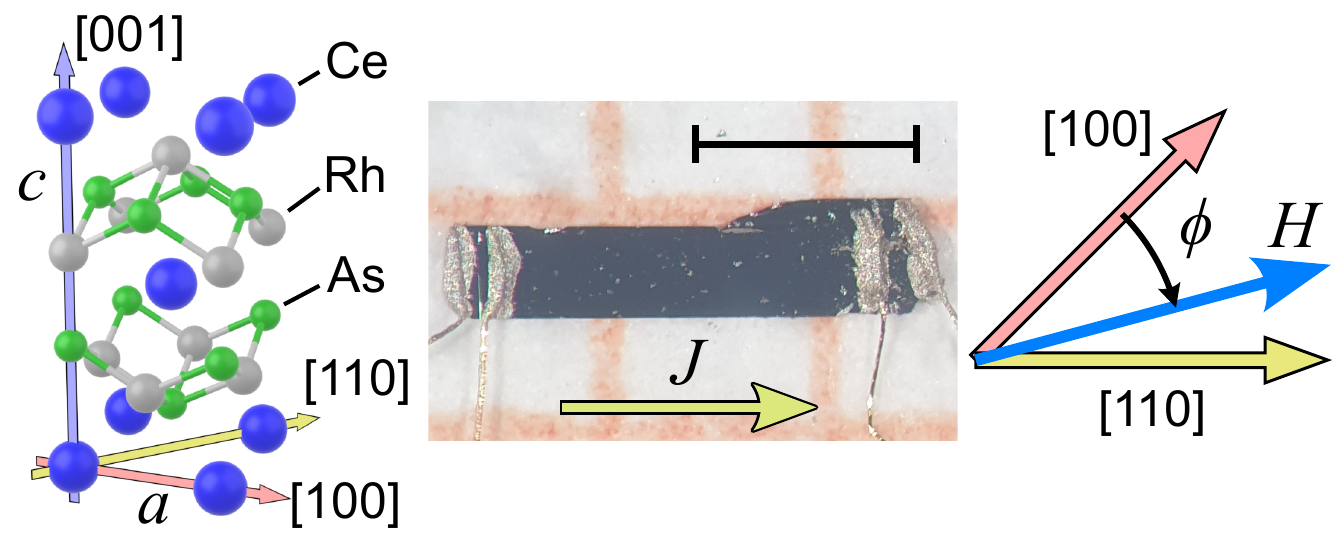}
    \caption{Left: tetragonal unit cell of \CRA. The crystallographic $a$ and $c$ axes as well as selected directions, labeled by Miller indices, are shown. Middle: photograph of the \CRA\ sample used for the resistivity measurements of this work. The yellow arrow indicates the direction of the current $J$. The scale bar marks a 1\,mm distance. Right: orientation of the sample in the middle panel. In this study, a magnetic field $H$ was applied perpendicular to the $c$ axis, at a varied angle $\phi$ with respect to the [100] orientation.}
    \label{fig:Intro}
\end{figure}

\CRA\ crystallises in a tetragonal non-symmorphic CaBe$_2$Ge$_2$-type structure (space group 129, $P4/nmm$)~\cite{madar1987}, shown in Fig.~\ref{fig:Intro}. When a magnetic field $\mu_0H$  (where $\mu_0$ is the vacuum permeability) is applied along the $c$ axis of the tetragonal unit cell, phase~I is suppressed similar to a typical antiferromagnetic (AFM) state and vanishes at about 7\,T~\cite{semeniuk2023,khanenko2025}. However, despite all the facts presented above, phase~I cannot be understood as a conventional AFM order because of the unusual robustness of \To\ against magnetic fields applied parallel to the basal plane of the lattice~\cite{hafner2022,mishra2022,chajewski2024}. Moreover, as can be seen in Fig.~\ref{fig:PhaseDiagrams}, upon exceeding a critical field ${\mu_0H_{\textrm{cr}}\lesssim9\rm \,T}$, \CRA\ undergoes a transition from phase~I into phase~II. The latter state is characterised by a strong continuous increase of \To\ with the field. A recent pulsed-field study reported a cascade of further phase transitions occurring past 20\,T, with a dome of electronic order stabilised up to 3.5\,K near 60\,T~\cite{blawat2025}.

Currently, the most promising explanation of the $H$-$T$ phase diagram of \CRA\ for $H\perp c$ suggests that phases I and II (collectively called the \To\ order) involve a quadrupolar ordering, which occurs in tandem with a conventional AFM state~\cite{hafner2022,schmidt2024}. At a first glance, there is a caveat with such a scenario: for tetragonal Ce-based systems, the crystal-electric-field (CEF) scheme consists of three ${J = 5/2}$ Kramer's doublets, whereas quadrupolar orders, in contrast, require a four-fold degenerate CEF ground state and are therefore only known to exist in certain cubic Ce-based crystals, rather than tetragonal ones~\cite{hanzawa1984}. In \CRA, however, the first excited state was found to be only $\sim30$\,K above the ground state~\cite{christovam2024}, and it has been shown theoretically that the two lowest-energy doublets can be mixed into a quasi-quartet by the Kondo coupling~\cite{hafner2022} and/or an in-plane ($a$ axis) magnetic field~\cite{schmidt2024}, thus enabling a quadrupolar order. Its presence in phases I and II can be investigated by studying how $T_{\textrm{0}}(H)$ varies upon changing the field orientation within the basal-plane of the lattice. An anisotropic behaviour, together with the field-induced enhancement of \To, would support a quadrupolar or higher-order multipolar ordering. Given the details of this anisotropy one may be able to quantify the relative strength of different multipole components and couplings involved and refine information on the CEF scheme of \CRA.

In this letter, we show that the phase diagram of \CRA\ for $H\perp c$ has a significant basal-plane anisotropy, as depicted in Fig.~\ref{fig:PhaseDiagrams}. The enhancement of \To\ by the field was found to be stronger for $H\parallel[110]$ than for $H\parallel[100]$, along with a higher value of the critical field \Hcr\ in the former case. The results cannot be readily captured by the existing model, motivating its extension as well as a re-examination of the CEF scheme of \CRA.

For our measurements, we used the latest generation of Bi-flux-grown \CRA\ crystals, which show a well-resolved and sharp heat capacity anomaly at \To~\cite{khanenko2025}. We probed electrical resistivity ($\rho$), which shows clear signatures of the \To\ transition~\cite{hafner2022,semeniuk2023}. The chosen sample was a thin elongated monocrystalline platelet with naturally formed extended flat facets normal to the $c$ axis and straight edges along the [110] direction and its equivalents, as determined via Laue X-ray diffraction (see Fig.~\ref{fig:Intro} for the annotated photograph of the sample). For a 4-point resistivity measurement, gold wires were connected to the sample via spot-welding and were additionally reinforced with the DuPont 4922 silver paint. The electrical current was supplied and the voltage was probed along the [110] direction. The residual resistivity ratio (RRR) of the sample was $\rho(300\,\mathrm{K})/\rho(0.5\,\mathrm{K})=3.0$. The crystal was mounted on a rotating stage of a dilution refrigerator, with the $c$ axis parallel to the rotation axis and perpendicular to the applied magnetic field. The uncertainty in the alignment was $\pm2\degree$. Measurements of heat capacity at high magnetic fields of up to 30\,T were conducted at the EMFL facility in Grenoble, using an AC calorimetry method (see the Methods section of Ref.~\cite{yang2024}) with a small, 0.12\,mg crystal that ensured a good sample homogeneity.

\begin{figure*}[t]
    \includegraphics[width=\textwidth]{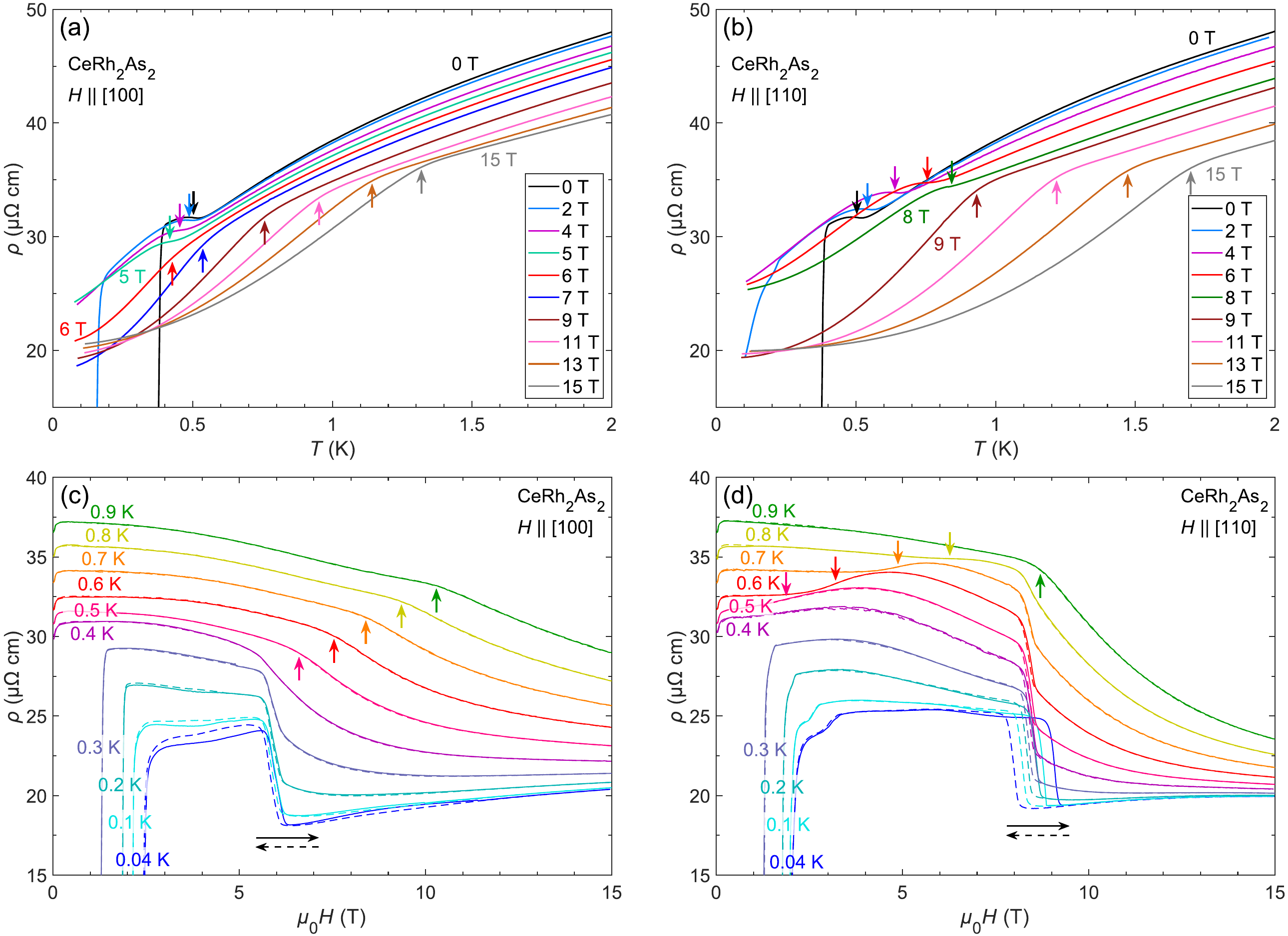}
    \caption{Electrical resistivity ($\rho$) of \CRA\ along the [110] crystallographic direction as a function of temperature $T$ (a,b) and magnetic field $\mu_{\textrm{0}}H$ (c,d) at fixed fields and temperatures, respectively. The field is applied along the [100] (a,c) and [110] (b,d) directions. The downward (upward) arrows mark the points of inflection (extremal curvature), defining the critical temperatures and fields of phase~I (II) of \CRA.}
    \label{fig:TSweepsHSweeps}
\end{figure*}

\textit{Results.}---We show the plots of $\rho(T)$ at fixed magnetic fields in Fig.~\ref{fig:TSweepsHSweeps}a,b. The transition into phase~I appears as an inflection of the curve, and we therefore used the minimum of $d\rho/dT$ as the definition of \To. Applying a magnetic field along the [100] direction (Fig.~\ref{fig:TSweepsHSweeps}a) initially causes a weak decrease of \To. Between 5 and 6\,T, the inflection of $\rho(T)$ is replaced by a downturn upon cooling, which is explained by phase~II setting in instead of phase~I. In the higher-field regime, we defined \To\ by the minimum of $d^2\rho/dT^2$ (the largest negative curvature). Above 6\,T, \To\ steadily increases with the field reaching 1.3\,K at 15\,T and up to 2.7\,K at 30\,T (see the heat capacity data below).

Applying a field along the [110] direction (Fig.~\ref{fig:TSweepsHSweeps}b) monotonically increases \To\ in the entire explored field range. The I--II transition occurs at a higher field, between 8 and 9\,T. Above \Hcr, \To\ increases with the field more rapidly than for $H\parallel[100]$, reaching 1.7\,K (2.8\,K) at 15\,T (29\,T).

In Fig.~\ref{fig:TSweepsHSweeps}c,d, we plot $\rho(H)$ at fixed temperatures for the two field orientations. At lower temperatures, the I--II transition is clearly identified as a step-like decrease of resistivity upon increasing the field (we defined \Hcr\ as the field at which the drop in $\rho(H)$ is the steepest). The first order character of the transition is apparent from the hysteresis. At temperatures high enough for the hysteresis to be negligible, \Hcr\ is equal to 5.8 and 8.4\,T for the [100] and [110] field directions, respectively.

We also used $\rho(H)$ curves for tracing the $T_{\textrm{0}}(H)$ line. Applying the same criteria as for \To\ (points of inflection and extremal curvature), we defined a critical field \Ho\ at which the system enters phases I and II (see Fig.~\ref{fig:TSweepsHSweeps}c,d).

A few visible features in our data were likely caused by extrinsic effects. The small dip of $\rho(H)$ below 0.2\,T indicates a presence of a small amount of an alien SC phase, possibly originating from the Bi flux. By extrapolating the $\rho(H)$ data down from above 0.2\,T, we obtained the best estimates of $\rho(H=0)$ at the respective temperatures. The zero-field $\rho(T)$ curve shown in Fig.~\ref{fig:TSweepsHSweeps}a,b has therefore been offset by a small positive value in order to correct for the influence of the alien phase. The existence of a surface state with the SC properties slightly different to those of the bulk may explain why the onset of the superconductivity in $\rho(H)$ for $H\parallel [110]$ has a double shoulder and is more gradual than for $H\parallel [100]$. The small differences between the upward and downward sweeps of $\rho(H)$ below 0.3\,K were likely caused by a slight temperature instability.

\begin{figure}
     \includegraphics[width=\columnwidth]{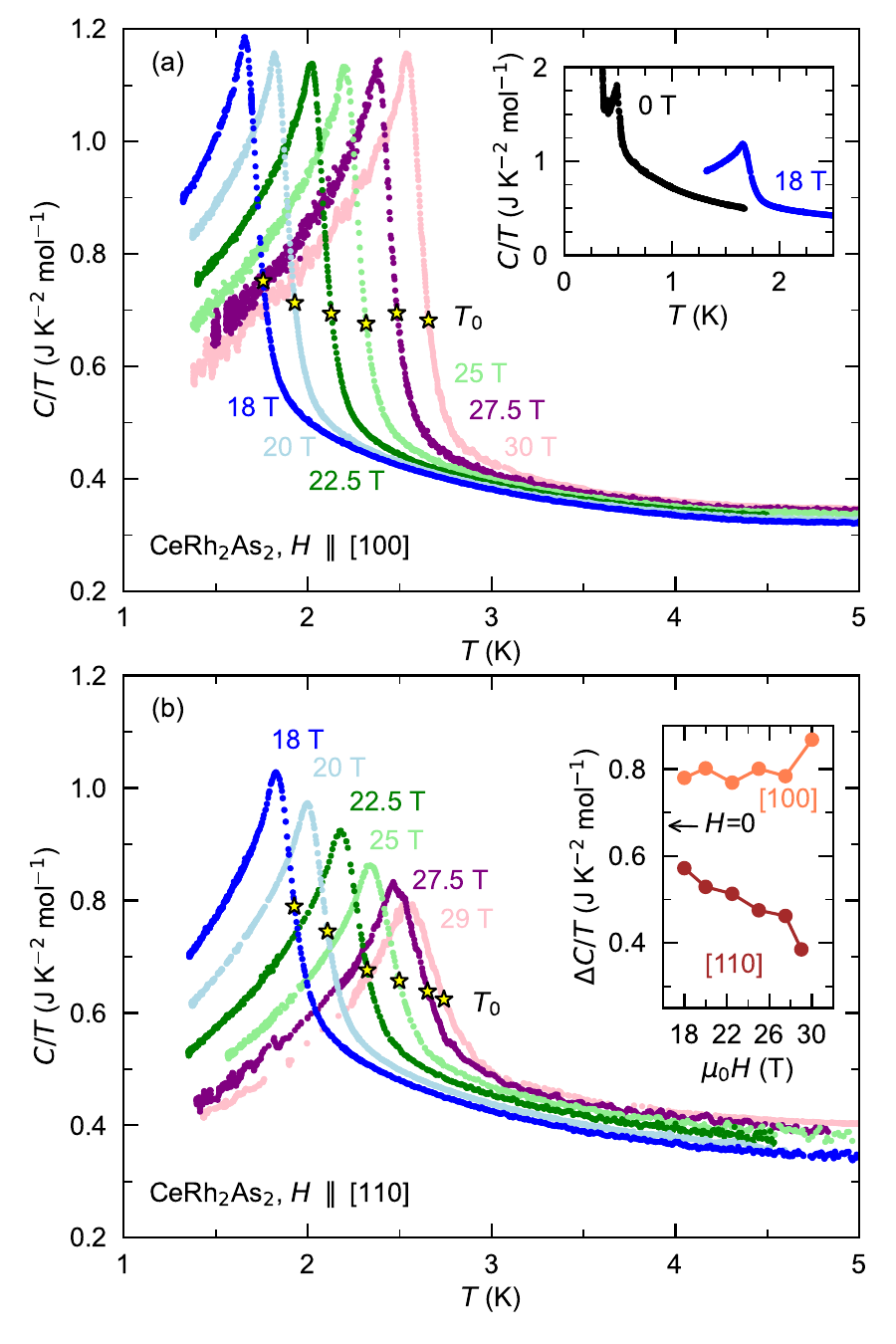}
     \caption{Heat capacity $C(T)$ of \CRA\ divided by temperature $T$ at high magnetic fields applied along the [100] (a) and [110] (b) crystallographic directions. The yellow stars mark the critical temperatures \To\ of phase~II, according to the equal entropy analysis. Panel~(a) inset provides a comparison with $C(T)$ at zero field. The inset in panel~(b) shows the corresponding heat capacity jump sizes $\Delta C/T$ at \To. The arrow marks the zero-field jump size}
     \label{fig:CTvsT}
\end{figure}

We supplement our resistivity study with high-field heat capacity data, shown in Fig.~\ref{fig:CTvsT}. The \To\ transition remains very pronounced up to 30\,T, evidencing that the anomaly remains a phase transition and does not become a crossover. Applying the equal-entropy condition (see Supplemental Material below), we analysed how the height of the ideal heat capacity jump $\Delta C/T$ varies with the field (inset of Fig.~\ref{fig:CTvsT}b). For $H\parallel[100]$, $\Delta C/T$ stays approximately constant. For $H\parallel[110]$, while \To\ is higher, $\Delta C/T$ is smaller and decreases with the field.

Using the data above, we compiled the field-temperature phase diagrams for $H\parallel[100]$ and $H\parallel[110]$, shown in Fig.~\ref{fig:PhaseDiagrams}. The anisotropy of the \To\ order for fields within the basal plane is the strongest in Phase~I, for which \To\ decreases with the [100]-oriented field, and increases with the [110]-oriented field. The high-field extensions of the phase boundaries based on the heat capacity data show a consistent anisotropy (Fig.~\ref{fig:PhaseDiagrams} inset). The high-field portion of the data also reveals a gradual flattening of the $T_{\textrm{0}}(H)$ line, in agreement with the dome-like shape observed in the pulsed-field study~\cite{blawat2025} and expected for a quadrupolar order~\cite{schmidt2024}.

\begin{figure}[t]
     \includegraphics[width=\columnwidth]{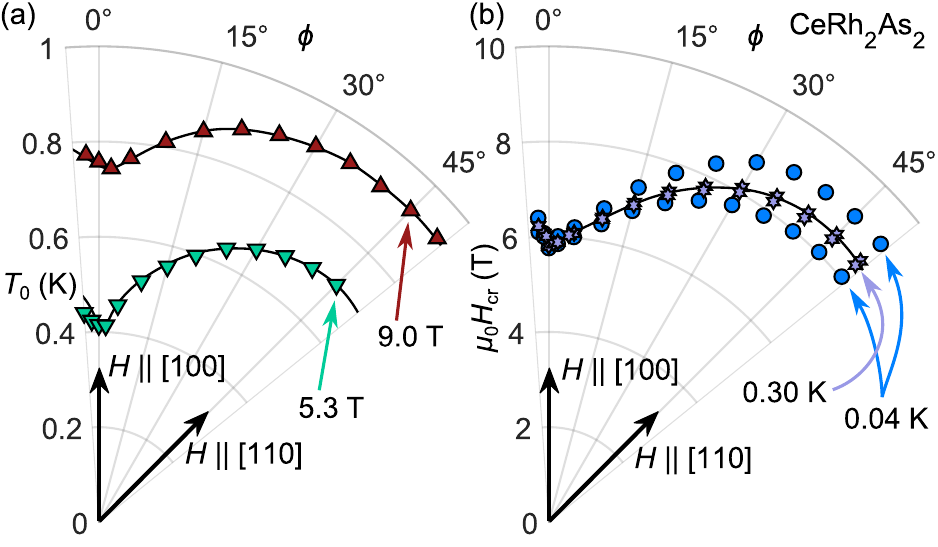}
     \caption{Magnetic-field-orientation dependence of the critical temperatures \To\ of phases I and II (a) and of the critical field \Hcr\ of the I--II transition (b) in \CRA. The angle $\phi$ measures the basal-plane orientation of the field with respect to the [100] crystallographic direction. Measurements of \To\ were done at 5.3 and 9.0\,T, when the system could only enter either phase~I or phase~II, respectively. The two values of \Hcr\ plotted for each angle correspond to the transitions encountered in the upward and downward field sweeps. The solid black lines are guides to the eye. The data used for extracting the \To\ and \Hcr\ values are provided in Supplemental Material.}
     \label{fig:T0andHcrAngleDependence}
\end{figure}

We further probed \To\ for a range of magnetic field orientations between the [100] and [110] directions (Fig.~\ref{fig:T0andHcrAngleDependence}a). The measurements were done with the field magnitudes fixed at 5.3 and 9.0\,T, so that only phase~I or phase~II, respectively, could set in for all field orientations. In both cases, \To\ is at its maximum at $H\parallel[110]$ and decreases monotonically with a narrow minimum at $H\parallel[100]$. The relative anisotropy is visibly stronger for phase~I ($A = T_0^{[110]}/T_0^{[100]}=1.75$ at 5.3\,T) compared to phase~II ($A=1.25$ at 9.0\,T). The field-angle dependence of \Hcr\ shows a similar picture (Fig.~\ref{fig:T0andHcrAngleDependence}b). The narrower hysteresis for $H\parallel[100]$ can be attributed to a significantly smaller value of \To\ at \Hcr.

\textit{Discussion.}---Our observed basal-plane anisotropy of the \To\ order in \CRA\ explains the apparent discrepancy in the previously published phase diagrams for in-plane fields. The phase diagram for $H\parallel[100]$ in Ref.~\cite{chajewski2024} agrees with our result, and namely with the decrease of \To\ with the field for $\mu_{\mathrm{0}}H<\mu_{\mathrm{0}}H_{\mathrm{cr}}^{[100]}=5.8$\,T. For the phase diagrams in Ref.~\cite{hafner2022} and Ref.~\cite{mishra2022}, the field was pointing along the [110] direction or close to it, for which \To\ increases with the field both below and above $\mu_{\mathrm{0}}H_{\mathrm{cr}}^{[110]}=8.4$\,T. The study in Ref.~\cite{hafner2022} found this value to be 9.0\,T, but given the strong sensitivity of the \To\ order to static crystalline defects, and their possible influence on the hysteresis, such a difference between different generations of crystals is not too surprising.

The pronounced basal-plane magnetic anisotropy of phases I and II cannot be understood in terms of a dipolar order alone. For a tetragonal system like \CRA\ with the point group $C_{4v}$ on the Ce site, the basal-plane components of the total magnetic dipole moment, $J_{x}$ and $J_{y}$, belong to the same two-dimensional irreducible representation $E$ ($\Gamma_{5}$ in Bethe notation). The only bilinear form they can take in the Hamiltonian, for it to remain invariant under a $C_{4}$ rotation, is $I_{\textrm{M}}(J_{i,x}J_{j,x}+J_{i,y}J_{j,y})$ ($I_{\textrm{M}}$ is the coupling constant; $i$, $j$ are the site indices). The exchange energy therefore depends on the relative orientation of the moments on two sites, rather than on the absolute ones, and a strictly dipolar order must respond to the in-plane field isotropically. For higher order multipoles, the moment components involved in the in-plane exchange generally do not belong to the same irreducible representation, making an anisotropic in-plane field response possible. This deduction is very much in line with the recent theoretical model that attempted to explain the unusual form of $T_{\textrm{0}}(H)$ for $H\parallel[100]$ with a field-induced antiferroquadrupolar (AFQ) order coupled to a conventional AFM order of the in-plane dipole~\cite{schmidt2024}.

\begin{figure}[t]
     \includegraphics[width=\columnwidth]{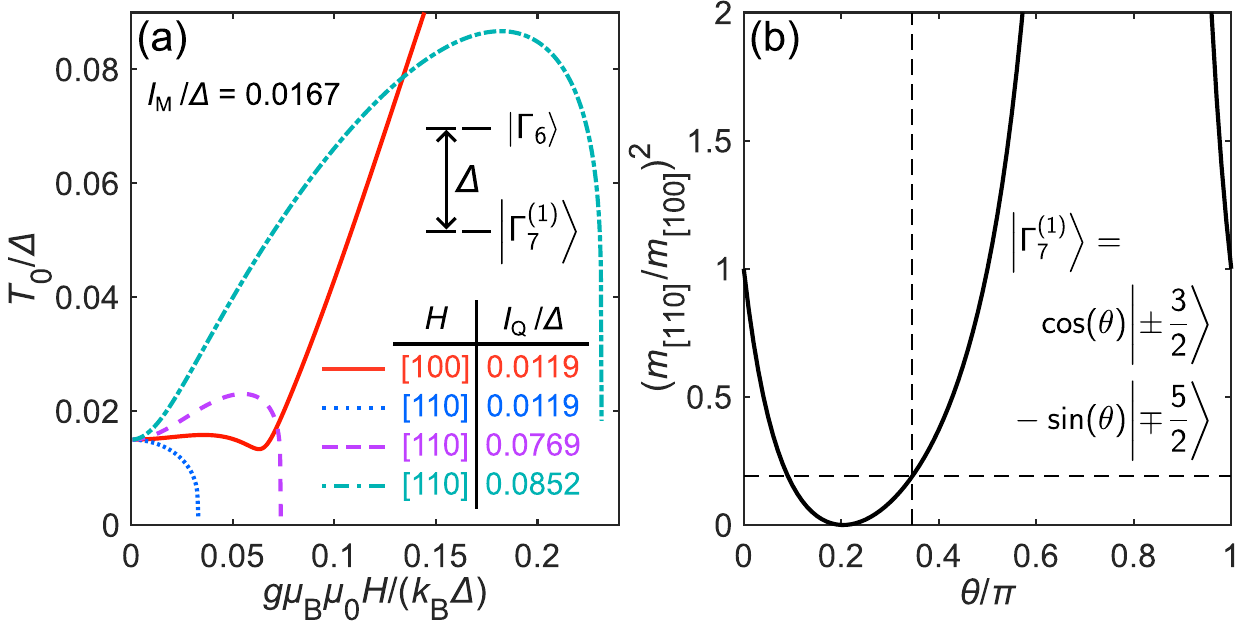}
     \caption{(a) Theoretical field ($H$) dependence of the critical temperature \To\ of \CRA, predicted by the magneto-quadrupolar ordering model~\cite{schmidt2024}. The parameter $\Delta$ is the zero-field splitting between the two lowest-energy crystal-electric-field doublets $\left|{\Gamma_{7}^{(1)}}\right\rangle$ and $\left|{\Gamma_{6}}\right\rangle$; $g$, $\mu_{\textrm{B}}$, $\mu_{\textrm{0}}$, and $k_{\textrm{B}}$ are the electron \textit{g} factor, Bohr magneton, vacuum permeability, and Boltzmann constant, respectively. The calculations were done for different orientations of $H$ and different values of the quadrupolar exchange coupling $I_Q$ (see the table). The in-plane isotropic magnetic exchange coupling $I_{\textrm{M}}$ was kept fixed. (b) Squared ratio of the matrix elements of the relevant quadrupole components for the two field directions as a function of the mixing angle $\theta$, which links the crystal-field states to the $\hat{J}_{z}$ basis.}
     \label{fig:theory}
\end{figure}

The essential ingredient of the model is the structure of the CEF wave functions, which is also crucial for understanding the pronounced basal-plane anisotropy in \CRA. Two of the three Kramers doublets transform according to the same irreducible representation~\cite{schmidt2024} and are therefore, in general, mixed, which is characterised by a (field-independent) mixing angle $\theta$. In Ref.~\cite{schmidt2024} it was shown for $H\parallel[100]$ that the physically relevant parameters are essentially given by $\xi_Q\propto m_{[100]}^2I_Q$ where $m_{[100]}=m_{[100]}(\theta,H)$ is the field-induced intra-doublet matrix element and $I_Q$ is the coupling of the $O_{xy}$ component of the quadrupole. This results in a phase boundary given by the solid red curve in Fig.~\ref{fig:theory}a, which captures the experimental findings surprisingly well, in a sense that it contains two regimes of distinguishable behaviour. At low fields, the AFM order parameter dominates over the AFQ one, and \To\ has a weaker field dependence. At higher fields, the dipolar order is suppressed, and the field-induced mixing of the first excited doublet into the CEF ground state generates and stabilises the AFQ order, causing the pronounced increase of \To. This outcome hints that phases I and II can be intuitively associated with a stronger and weaker role of the dipolar interaction, respectively (and vice versa for the quadrupolar interaction).

An obvious shortcoming of the model is its prediction of fundamentally the same ordered state in the entire domain of $T_{\textrm{0}}(H)$, while the experiment shows that phases I and II are distinct competing orders. Realising the first order transition at \Hcr\ clearly requires a more complex description, including, for instance, magnetoelastic coupling or the Kondo effect. The fact that \CRA\ is a metal is strongly suggestive of a density wave picture of the \To\ order~\cite{hafner2022}. The distinct forms of the resistivity anomalies of phases I and II could reflect different Fermi surface reconstructions, with the \Hcr\ transition corresponding to a change of the nesting vector. However, given that the electronic structure of the compound is inevitably linked to the crystal-field environment of the Ce site, it is not unexpected that a description based purely on local properties of $f$ electrons can provide insights into the physics of the system (see Ref.~\cite{thalmeier2022}, for instance).

It is natural to extend the model to the ${H\parallel[110]}$ case. The field then induces matrix elements $m_{[110]}=m_{[110]}(\theta,H)$ within the ground-state doublet of the $O_{x^2-y^2}$ quadrupole component, in a similar manner as for $O_{xy}$ in the ${H\parallel[100]}$ case, but with a very different $\theta$ dependence. The results are summarised in Fig.~\ref{fig:theory}: panel (a) shows the \To\ phase boundaries for ${H\parallel[110]}$ for three different choices of the AFQ coupling $I_Q(O_{x^2-y^2})$. The dotted blue curve corresponds to $I_Q(O_{x^2-y^2})=I_Q(O_{xy})$, for the dashed purple and dash-dot teal curves we chose $I_Q(O_{x^2-y^2})\gg I_Q(O_{xy})$. All other parameters, namely the mixing angle $\theta$, the splitting $\Delta$ of the two low-lying doublets, and the isotropic in-plane AFM exchange $I_{\textrm{M}}$ are the same as in Ref.~\cite{schmidt2024}. For $I_Q(O_{x^2-y^2})=I_Q(O_{xy})$, the \To\ order shows no field-induced enhancement for ${H\parallel[110]}$ and is strongly suppressed compared to the ${H\parallel[100]}$ case. This theoretical result is in stark contrast with our experimental findings, which can be attributed to two likely causes:

1. The matrix elements $m_{[110]}$ of the $O_{x^2-y^2}$ quadrupole component are strongly suppressed compared to $m_{[100]}$ for the $O_{xy}$ component. For fields $h\ll\Delta$ this is illustrated in Fig.~\ref{fig:theory}b where the vertical dashed line marks the mixing angle taken from Ref.~\cite{hafner2022}, and the horizontal dashed line indicates the corresponding ratio $(m_{[110]}/m_{[100]})^2\approx0.29$. As seen in Fig.~\ref{fig:theory}a, the influence of the matrix elements can be at least partially ``compensated'' by increasing the quadrupolar coupling strength $I_Q(O_{x^2-y^2})$. Note, that in the somewhat artificial case $\theta=0$ (no mixing), $(m_{[110]}/m_{[100]})^2=1$, and the phase diagram is fully isotropic with respect to the basal-plane field direction.

2. The in-plane quadrupolar components $O_{xy}$ and $O_{x^2-y^2}$ transform according to different irreducible representations, unlike the in-plane dipole components. Thus, there is no a priori reason why the combined AFM+AFQ phase boundary should be identical or even similar for different field directions. This is also reflected in the structure of the field-induced matrices of $O_{xy}$ and $O_{x^2-y^2}$. The former contains pseudo-van-Vleck intra-doublet elements~\cite{schmidt2024} while the latter contains pseudo-Curie type intra-doublet elements (see Supplemental Material below).

The comparison of the model with the experimentally observed basal-plane anisotropy of $T_{\textrm{0}}(H)$, and particularly the phase diagram for $H\parallel[110]$, suggests that a modification to the currently assumed CEF scheme of \CRA\ is necessary in order to better capture the phenomenology of the system.

\textit{Conclusions.}---In this work, we exposed the pronounced magnetic basal-plane anisotropy of the ordered states of \CRA, known as phases I and II. The contrast between the fields along the [100] and [110] crystallographic directions is particularly strong for phase~I, with the critical temperature \To\ decreasing and increasing with the field, respectively. This result supports the idea of the \To\ transition involving quadrupolar (or even higher-order multipolar) degrees of freedom, setting \CRA\ apart from all other Ce-based tetragonal heavy-fermion systems.

\begin{acknowledgments}
We thank P.~Thalmeier, G.~Knebel, A.~Pourret, S. Ruet, M.~Brando, and G.~Zwicknagl for fruitful discussions. The work was supported by the Deutsche Forschungsgemeinschaft (DFG) through CRC1143 (Project No. 247310070), the Würzburg-Dresden Cluster of Excellence on Complexity and Topology in Quantum Matter—ct.qmat (EXC 2147,
Project ID 390858490), and the European Research Council (ERC) grant Ixtreme, GA 101125759. The authors also acknowledge the support of the LNCMI-CNRS, a member of the European Magnetic Field Laboratory (EMFL).
\end{acknowledgments}
\vspace{\fill}
\clearpage


\begin{center}
\textbf{\large Supplemental Material}
\end{center}

\setcounter{equation}{0}
\setcounter{figure}{0}
\setcounter{table}{0}
\makeatletter
\renewcommand{\theequation}{S\arabic{equation}}
\renewcommand{\thefigure}{S\arabic{figure}}
\renewcommand{\thetable}{S\arabic{table}}

\section{Determination of $T_{0}$ from the heat capacity data via the equal-entropy analysis}

\begin{figure}[b!]
     \includegraphics[width=\columnwidth]{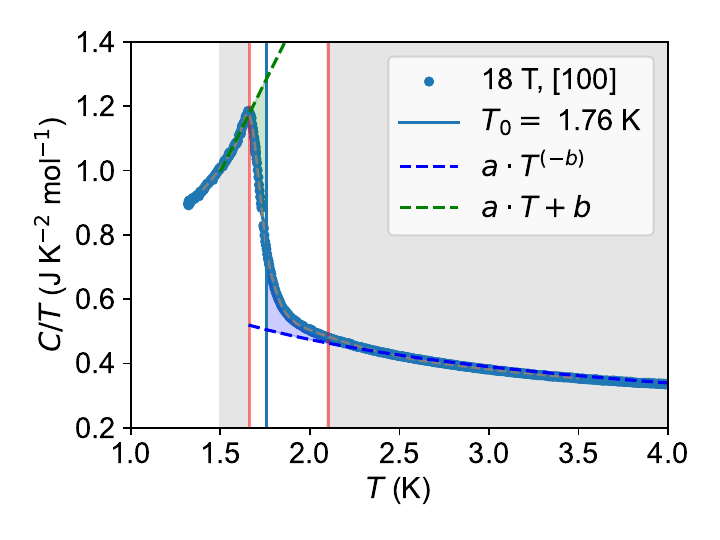}
     \caption{Illustration of the equal entropy method for determining the critical temperature \To\ of \CRA\ using the temperature ($T$) dependence of heat capacity divided by temperature ($C/T$). The example uses the data for 18\,T magnetic field in the [100] direction (blue markers). The broadened heat capacity anomaly at the second-order phase transition is bounded by the two vertical red lines. Part of the curve located in the grey-shaded above (below) the anomaly is extrapolated across the anomaly using an power-law (linear) function. The extrapolations are shown in dashed lines and are supposed to reflect the behaviour of $C/T$ in the ordered and unordered states for an ideal crystal. The parameters $a$ and $b$ are distinct for each extrapolation. The value of \To\ (blue vertical line) is chosen such that the area shaded in green is matched by the area shaded in blue.}
     \label{sfig:equal_entropy}
\end{figure}

Fig.~\ref{sfig:equal_entropy} illustrates the method for determining the critical temperature \To\ from the heat capacity data (Fig.~3 of the main paper), using the equal entropy construction. For an ideal heat capacity anomaly, the critical temperature of a second-order phase transition corresponds to the point where $C$ changes discontinuously. In reality, various inhomogeneities throughout the sample broaden the anomaly, but is sensible to assume that if we compare two temperatures, clearly below and above the range spanned by the anomaly, then the change in entropy between these two temperatures  (area bounded by $C(T)/T$) is effectively the same as for an ideal crystal, since they correspond to the sample being completely in the ordered and disordered state, respectively. In order to determine \To, one needs to extrapolate the $C(T)/T$ curve for the ordered and unordered states towards the transition, and find such a position for the jump between them, that results in both the ideal and experimental curves resulting in the same entropy change across the anomaly.

\section{Resistivity data for the angular dependences of $T_{0}$ and $H_{\textrm{cr}}$}

\begin{figure}[b!]
     \includegraphics[width=\columnwidth]{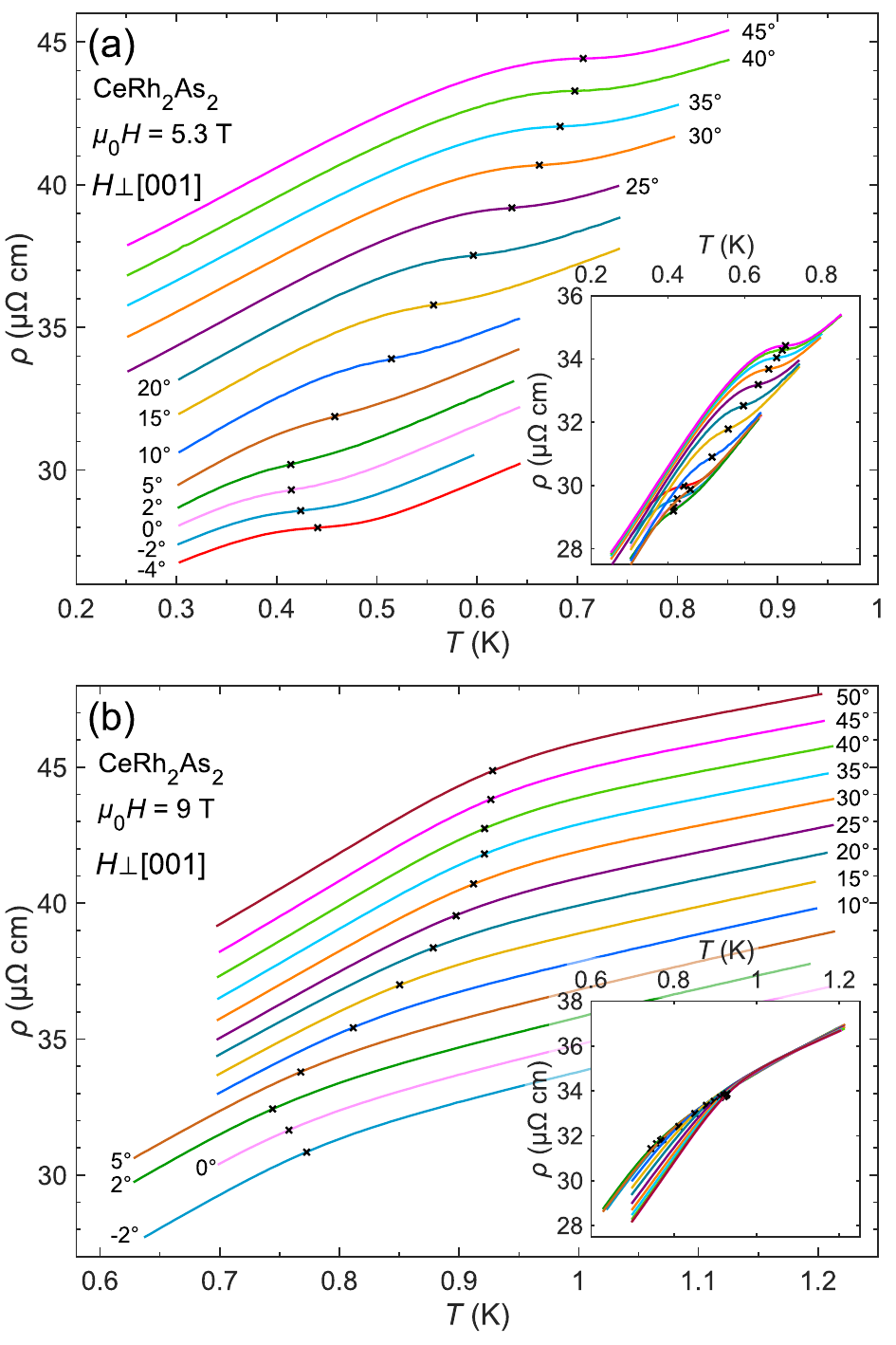}
     \caption{Electrical resistivity ($\rho$) of \CRA\ as a function of temperature ($T$) near the critical temperature \To (black crosses), for a magnetic field ($H$) applied along different directions in the basal plane. The angle of 0$^\circ$ corresponds to the field along the [100] crystallographic orientation. Resistivity was probed with the current flowing along the [110] direction. The magnitude of the field was set to 5.3\,T (a) and 9.0\,T (b), such that the system only enters Phase~I or Phase~II, respectively. The curves are offset in steps of 1\,$\micro\Omega$\,cm, with respect to the 0$^\circ$ curve. The insets show the same data without the offsets.}
     \label{sfig:rho_v_T_angles}
\end{figure}

\begin{figure}
     \includegraphics[width=\columnwidth]{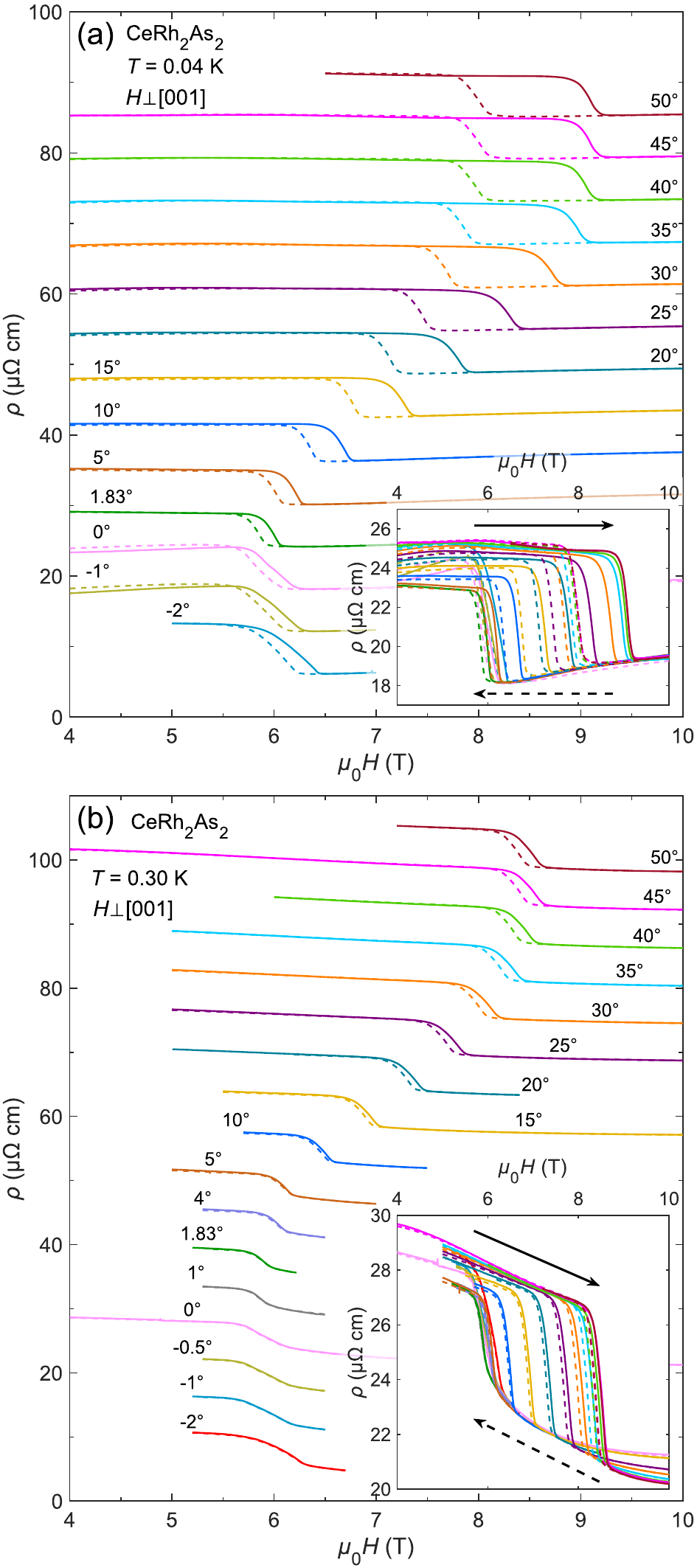}
     \caption{Electrical resistivity ($\rho$) of \CRA\ as a function of magnetic field ($H$) across the critical field \Hcr\ of the Phase~I-Phase~II transition. The field was applied along different directions in the basal plane. The angle of 0$^\circ$ corresponds to the field along the [100] crystallographic orientation. Resistivity was probed with the current flowing along the [110] direction. The data were collected at different temperatures ($T$): 0.04\,K (a) and 0.30\,K (b). The curves are offset in steps of 6\,$\micro\Omega$\,cm, with respect to the 0$^\circ$ curve. The insets show the same data without the offsets.}
     \label{sfig:rho_v_H_angles}
\end{figure}

In Fig.~\ref{sfig:rho_v_T_angles} and Fig.~\ref{sfig:rho_v_H_angles}, we show the field and temperature sweeps of resistivity for the field applied at different angles in the basal plane. These data provided the values of \To\ and \Hcr\ for the polar plots in Fig.~5 of the main paper.

In the plots of $\rho(T)$ in the insets of Fig.~\ref{sfig:rho_v_T_angles}, one can see that above \To\ all the curves collapse onto the same line, and the basal plane anisotropy develops only below \To. For the $\rho(H)$ plots, resistivity appears to be rather isotropic in Phase~II at 0.04\,K (Fig.~\ref{sfig:rho_v_H_angles}a inset), but at 0.3\,K resistivity varies with the angle both below and above \Hcr\ (Fig.~\ref{sfig:rho_v_H_angles}b inset).

\section{The $O_{x^2-y^2}$ quadrupole component}

\subsection{The single-ion Hamiltonian}

In this section, we closely follow the derivations given in Ref.~\cite{schmidt2024}. Now we assume the applied magnetic field to be oriented in the $[110]$ direction. First of all, let's diagonalise the single-ion Hamiltonian
\begin{eqnarray}
	{\cal H}
	&=&
	H_\text{CEF} + H_\text{Zeeman},
	\label{eqn:multipoles:hcefpluszeeman}
	\\[\baselineskip]
	H_\text{CEF}
	&=&
	B_2^0O_2^0+B_4^0O_4^0+B_4^4O_4^4,
	\label{eqn:multipoles:hcef}
	\\
	H_\text{Zeeman}
	&=&
	-\underbrace{g_j\mu_\text B\mu_0 H}_h\frac1{\sqrt2}\left(J_x+J_y\right)
	=:-hJ_{xy}.
\end{eqnarray}
Here, $h$ is the sum of the external field $H_0$ and the induced molecular field along $[110]$. The spectrum of $H_\text{CEF}$ is comprised of three Kramers doublets, the two low-lying of which form a quasi-quartet which we discuss from now on. We apply to it a series of unitary transformations, namely first an intra-doublet state transformation and then an inter-doublet state rotation, the latter characterised by two angles $\alpha_j$, $j=1,2$. Bringing $\cal H$ into diagonal form then requires
\begin{eqnarray}
	\tan(2\alpha_j)
	&=&
	(-)^{\hat j}\frac{2hm_{a}'}{-\Delta+(-)^jh(m_{a_1}-m_{a_2})}
	\label{eqn:multipoles:alpha}
	\\
	\Rightarrow\alpha_j
	&\approx&
	(-)^j\frac{hm_{a}'}\Delta,
	\quad
	\frac h\Delta\ll1
	\label{eqn:multipoles:smallalpha}
\end{eqnarray}
with $\hat j:=3-j$ and $\alpha_j\ne\pi/4+n\pi$, $n\in\mathbb{Z}$. The constants in Eq.~(\ref{eqn:multipoles:alpha}) are
\begin{eqnarray*}
	m_{a_1} &:=& -\frac{\sqrt5}2\sin(2\theta), \\
	m_{a_2} &:=& \frac32, \\
	m_a' &:=& -\sqrt2\sin(\theta)
\end{eqnarray*}
with $\theta$ being the crystal-field mixing angle related to the crystal-field parameters like
\begin{equation}
	\tan(2\theta)
	=
	\frac{-2\sqrt5B_4^4}{B_2^0+20B_4^0}
\end{equation}
and $\theta\ne\pi/4+n\pi$, $n\in\mathbb{Z}$.

\subsection{The quadrupole component}

The $H=0$ bare quadrupole operator expressed in terms of the crystal-field (CEF) basis states has the form
\begin{equation}
	O_{x^2-y^2}(\theta)
	=
	\left(
	\begin{array}{cccc|cc}
	 0 & 0 & m_{Q_3}' & 0 & 0 & 0 \\
	 0 & 0 & 0 & m_{Q_3}' & 0 & 0 \\
	 m_{Q_3}' & 0 & 0 & 0 & m_{Q_4}' & 0 \\
	 0 & m_{Q_3}' & 0 & 0 & 0 & m_{Q_4}' \\
	 \hline
	 0 & 0 & m_{Q_4}' & 0 & 0 & 0 \\
	 0 & 0 & 0 & m_{Q_4}' & 0 & 0 \\
	\end{array}
	\right)
\end{equation}
where the first four rows and columns denote the quasi-quartet. The matrix elements are given by
\begin{eqnarray*}
	m_{Q_3}'
	&:=&
	\sqrt2\left(\sqrt5\cos(\theta)-3\sin(\theta)\right),
	\\
	m_{Q_4}'
	&:=&
	\sqrt2\left(3\cos(\theta)+\sqrt5\sin(\theta)\right).
\end{eqnarray*}
Applying the same unitary transformations which bring the Hamiltonian~(\ref{eqn:multipoles:hcefpluszeeman}) restricted to the quasi-quartet into diagonal form, we eventually obtain
\begin{equation}
	O_{x^2-y^2}
	\to
	\left(
	\begin{array}{cc|cc}
	-\tilde M_{Q_3}^{1\text s} & 0 & \tilde M_{Q_3}^{1\text c} & 0 \\
	0 & -\tilde M_{Q_3}^{2\text s} & 0 & \tilde M_{Q_3}^{2\text c} \\
	\hline
	\tilde M_{Q_3}^{1\text c} & 0 & \tilde M_{Q_3}^{1\text s} & 0 \\
	0 & \tilde M_{Q_3}^{2\text c} & 0 & \tilde M_{Q_3}^{2\text s} \\
	\end{array}
	\right).
	\label{eqn:multipoles:tildeox2y2}
\end{equation}
with field-induced ($\alpha_j\ne0$) matrix elements
\begin{equation}
	\tilde M_{Q_3}^{j\eta}
	:=
	m_{Q_3}'\left\{
	\begin{array}{ll}
		\sin(2\alpha_j), & \eta=\text s \\
		\cos(2\alpha_j), & \eta=\text c
	\end{array}
	\right.
\end{equation}
depending both on field and CEF mixing angle. We stress that only for field $h\ne0$ we obtain finite intra-doublet matrix elements $\tilde M_{Q_3}^{j\text s}\ne0$. In the main text, we use the notation $m_{[110]}=|\tilde M_{Q_3}^{j\text s}|$.

\bibliography{References}

\end{document}